\documentclass[11pt,onecolumn,floatfix,superscriptaddress,nofootinbib]{revtex4-1}

\usepackage[english]{babel}
\usepackage{amsmath}
\usepackage{graphicx}                   
\usepackage{epsfig}
\usepackage{amsmath,amssymb}            
\usepackage{hyperref}                   
\usepackage{epstopdf}
\usepackage{placeins}
\usepackage{upgreek}
\usepackage{slashed}
\usepackage{color}
\usepackage{footmisc}
\usepackage{footnote}
\definecolor{orange}{rgb}{1,0.5,0}
\definecolor{brown}{rgb}{0.65, 0.16, 0.16}
\definecolor{phlox}{rgb}{0.87, 0.0, 1.0}

\graphicspath{{figs/}}                  
\usepackage{ulem,xcolor}

\def\al{\alpha}

\def\ep{\epsilon}

\def\ka{\kappa}
\def\la{\lambda}

\def\mn{{\mu\nu}}
\def\cl{{\cal L}}

\def\prt{\partial}

\def\half{{\textstyle{1\over 2}}}
\def\frac#1#2{{\textstyle{{#1}\over {#2}}}}

\def\lsim{\mathrel{\rlap{\lower4pt\hbox{\hskip1pt$\sim$}}
    \raise1pt\hbox{$<$}}}
\def\gsim{\mathrel{\rlap{\lower4pt\hbox{\hskip1pt$\sim$}}
    \raise1pt\hbox{$>$}}}
\def\sqr#1#2{{\vcenter{\vbox{\hrule height.#2pt
         \hbox{\vrule width.#2pt height#1pt \kern#1pt
         \vrule width.#2pt}
         \hrule height.#2pt}}}}

\newcommand{\beq}{\begin{equation}}
\newcommand{\eeq}{\end{equation}}
\newcommand{\bea}{\begin{eqnarray}}
\newcommand{\eea}{\end{eqnarray}}

\def\kaf{k_{AF}}
\def\kf{k_{F}}

\begin{document}

    \title{The optical analogy between a Lorentz-violating cosmos and a Magneto-Electric medium}
\author{Iman Motie}
\author{Brahim Lamine}
\author{Alain Blanchard}
\affiliation{Université de Toulouse, UPS-OMP, IRAP, F-31400 Toulouse, France }	
\author{Rémy Battesti}
\author{Carlo Rizzo}
\email{carlo.rizzo@lncmi.cnrs.fr}
\affiliation{Université de Toulouse, LNCMI UPR CNRS 3228 (UGA, INSA-T, EMFL), F-31400 Toulouse Cedex, France}

        
\begin{abstract}
   
The goal of our study is to investigate the effects of Lorentz symmetry violation by examining the behavior of photons within the framework of the Standard Model Extension (SME). 
We show that, from an optical point of view, the Lorentz-violating cosmos is analogous to a Magneto-Electric medium like the quantum vacuum in the presence of the magnetic and electric field. This analogy provides a formulation for exploring experimental evidence of Lorentz symmetry violation in the propagation of light in a vacuum, in particular in the case of a radiation background observed like the CMB. 
 
\end{abstract}


\maketitle

\section{Introduction}
Lorentz symmetry is a fundamental concept in physics, forming the foundation of both special and general relativity. It asserts that the laws of physics remain unchanged under Lorentz transformations, meaning that no specific inertial frame of reference is preferred. This principle ensures that the speed of light is constant for all observers, regardless of their motion, and it has profoundly shaped our understanding of space, time, and the interactions of particles. Lorentz symmetry underpins major theories such as quantum field theory, general relativity, and the Standard Model of particle physics, playing a crucial role in connecting the behavior of particles and forces at microscopic scales with the dynamics of the universe on a cosmic scale (see e.g. \cite{Tasson2014}).

Because of its central importance, the possibility of Lorentz symmetry breaking has become an active area of research. Under extreme conditions, such as very high-energy scales close to the Planck energy or in regions with intense gravitational fields, deviations from Lorentz symmetry might occur. Exploring these potential violations could reveal new physics beyond the Standard Model. The Standard Model Extension (SME) \cite{Colladay1998} \cite{Colladay:1996iz}, provides a framework to systematically study such scenarios. By introducing additional terms to the Standard Model Lagrangian, the SME allows physicists to investigate how violations of Lorentz symmetry could affect various physical phenomena.

One particularly interesting area of study is how Lorentz symmetry breaking affects the propagation of photons. In the SME framework, the behavior of photons can be altered, leading to effects such as changes in their speed or polarization. These modifications can be interpreted as similar to the behavior of photons traveling through a medium. A notable example is birefringence, i.e. the dependence of the velocity of light on the light polarization circular mode, in the case of circular birefringence, or the direction of polarization in the case of linear birefringence, due to Lorentz-symmetry violations. \textcolor{black}{Moreover, specific optical effects known in material sciences, but existing also in a vacuum following Quantum ElectroDynamics (QED) \cite{Battesti2013}, such as Kerr \cite{Kerr1875}, Cotton-Mouton \cite{Cotton}, Jones \cite{Roth2000}, and Magneto-Electric \cite{Roth2002} birefringences, provide useful analogues for understanding these phenomena in the context of SME \cite{Colladay1998}. These effects involve anisotropic responses of media to electric and magnetic fields, closely resembling the theoretical modifications of photon propagation in a Lorentz-violating vacuum. This connection between symmetry breaking and measurable effects opens up opportunities for experimental tests.}

Astrophysical observations are especially useful for investigating these effects. Light from distant cosmic sources, such as gamma-ray bursts or active galactic nuclei, travels vast distances before reaching us. Any violation of Lorentz symmetry along the way would leave imprints on the light, such as shifts in its speed or changes in polarization, that may be detectable. These small deviations could be observed with advanced instruments, providing evidence for or against Lorentz symmetry breaking. Additionally, studying light in extreme environments, such as near black holes or during high-energy events, offers a promising way to detect these effects (see e.g. \cite{Kislat2018}). Terrestrial experiments have also been conducted to detect such a violation using a large panel of systems going from fundamental particles like neutrinos (see e.g.\cite{Díaz2016}) to bound systems or light in resonant cavities (see e.g.\cite{Russell2005}). However, no evidence of a Lorentz violation has yet been found. An extensive review of the current limits on SME parameters can be found in \cite{Kostelecky2025}.
\textcolor{black}{In particular, recent studies focusing on cosmic microwave background (CMB) polarization have set tighter constraints on SME coefficients (see e.g.\cite{iman, Leon_2016, shao, Pogosian2019, feri, Caloni_2023} and references within), demonstrating the increasing sensitivity of observations and highlighting the need for new theoretical predictions that could be tested with future experiments.} 

An intriguing parallel can be drawn between Lorentz-violating models and the behavior of photons in Magneto-Electric media in which electric and magnetic properties are intrinsically coupled. This kind of anisotropic media exhibit special features in light propagation \cite{O'Dell1962} that can be mimicked in a Lorentz violating cosmos if certain SME coefficients are nonzero. Consequently, Magneto-Electric materials serve as analog models for testing SME theoretical predictions on optical Lorentz-violating effects. 
As far as light propagation is concerned, a quantum vacuum in the presence of electromagnetic fields behaves as well as a Magneto-Electric crystal because of its electric and magnetic response to field perturbation \cite{Battesti2013} following the prediction of Quantum ElectroDynamics (QED).

The analogy between light propagation in the framework of SME and in macroscopic media has been recognized by Colladay and Kostelecky since their 1998 seminal paper \cite{Colladay1998}, but not fully exploited until now.

\textcolor{black}{The purpose of our paper is to study this analogy in  details. The interest for such studies has been renewed recently by detection of possible cosmic birefregence in the CMB \cite{Komatsu}. It} is organized as follows :

In Section \ref{review}, we briefly overview the pure photon sector of the SME. Section \ref{analogy} draws an analogy between Lorentz-violating electrodynamics in a vacuum and homogeneous anisotropic media, discussing magneto-electric media and the quantum vacuum as an effective magneto-electric medium. We analyze optical activity and the effects of Lorentz violation on the CPT-even term and birefringence. Section \ref{secconc} offers concluding remarks. Detailed calculations on photon birefringence can be found in Appendix \ref{appe}.


 \section{ Pure Photon Sector of the SME }\label{review}
  
The photon sector of the SME represents one of the most intriguing components for exploring Lorentz symmetry breaking \cite{Colladay1998}. Photons, being massless particles and carriers of the electromagnetic force, play a crucial role in probing the fundamental properties of space-time. The SME modifies the conventional Maxwell Lagrangian by introducing additional terms that account for possible violations of Lorentz symmetry. These terms, parameterized by coefficients for Lorentz violation, allow for deviations from the standard behavior of electromagnetic waves.

In the SME framework, the modified Lagrangian for photons can be expressed as:
\begin{eqnarray}
\mathcal{L}&=& =  -\frac 1 4 F_{\mu\nu}F^{\mu\nu}
+\frac 1 2 (\kaf)^\rho\ep_{\rho\la\mu\nu}A^\la F^{\mu\nu}
- \frac 1 4 (k_F)_{\rho\lambda\mu\nu} F^{\rho\la}F^{\mu\nu} ,
\label{lagrangian}
\end{eqnarray}
where $F_\mn \equiv \prt_\mu A_\nu -\prt_\nu A_\mu$ is the electromagnetic field strength tensor, $A_\mu$ is the electromagnetic potential, $(k_F)_{\rho\lambda\mu\nu}$ 
 represents the dimensionless coefficients for CPT-even Lorentz violation. It will be useful to decompose the coefficient $k_F$
into a tensor with 10 independent components analogous to the Weyl tensor in general relativity and one with 9 components analogous
to the trace-free Ricci tensor with the following symmetries of the Riemann tensor
 \beq \label{symm}
 \:(k_{F})_{\mu\nu\alpha\beta}=
 -\:(k_{F})_{\nu\mu\alpha\beta}=-\:(k_{F})_{\mu\nu\beta\alpha}=\:(k_{F})_{\nu\mu\beta\alpha}=\:(k_{F})_{\alpha\beta\mu\nu}.\eeq
So it contains 19 independent real components.
$(\kaf)^\rho$ has dimensions of mass and represents the coefficients for CPT-odd Lorentz violation \cite{Colladay:1996iz, Colladay1998}. 

The CPT-odd term has been the subject of considerable interest in the literature \cite{jk}. This term presents some challenges, as it contributes negatively to the canonical energy, potentially leading to instability. A constructive approach to address this issue is to set the coefficient to zero, $(\kaf)^\ka = 0$. This solution aligns well with theoretical considerations related to radiative corrections in the standard model extension and is strongly supported by experimental results. Notably, strict constraints on $\kaf$ have been established through the study of radiation polarization from distant radio galaxies \cite{cfj}.

In contrast, there is an opportunity for further exploration with respect to the CPT-even coefficient $\kf$. Theoretical analyses suggest that it provides positive contributions to the canonical energy and is radiatively induced from the fermion sector within the standard model extension \cite{Colladay:1996iz, Colladay1998}. Recent advances have led to constraints on some components of $\kf$, obtained from optical spectropolarimetry of distant cosmological sources \cite{Komatsu}. 

Regarding  $\ka$ matrices  and assuming that $(\kaf)^\al = 0$, the Lagrangian (\ref{lagrangian}) can be written as \cite{sig},
\bea
\cl&=&\half(\epsilon_0\textbf{E} ^2-\frac{\textbf{B} ^2}{\mu_0})
+\frac{1}{2} \epsilon_0\textbf{E}\cdot(\ka_{DE})\cdot\textbf{E}
-\half\frac{\textbf{B}}{\mu_0}\cdot(\ka_{HB})\cdot\textbf{B}
+\sqrt{\frac{\epsilon_0}{\mu_0}}\textbf{E}\cdot(\ka_{DB})\cdot\textbf{B}.
\label{lagkappa}
\eea
where in this equation $\epsilon_0$ and $\mu_0$ are the vacuum permittivity and permeability, respectively, $\textbf  E$ and $\textbf B$ are the electric and magnetic fields obtained from solving the modified Maxwell equations.

The $3 \times 3$ matrices $\ka_{DE}$, $\ka_{HB}$, $\ka_{DB}$,
and $\ka_{HE}$ are defined by \cite{sig}
\bea
(\ka_{DE})^{jk} &=& -2 (\kf)^{0j0k}, 
\nonumber \\
(\ka_{HB})^{jk} &=& \half \ep^{jpq} \ep^{krs} (\kf)^{pqrs}, 
\nonumber \\
(\ka_{DB})^{jk} &=& -(\ka_{HE})^{kj} = (\kf)^{0jpq}\ep^{kpq}.
\label{kappas}
\eea

Using Eqs.~(\ref{lagkappa}), in the context of Lorentz violation, the polarization and magnetization will also depend on the Lorentz-violating coefficients as follows:
\begin{eqnarray}
    {P}_i&=&\frac{\partial \cl}{\partial E_i}-\epsilon_0 E_i=\half\epsilon_0 (\ka_{DE})_{ik}{E}^k+\half\epsilon_0 {E}^j(\ka_{DE})_{ji}\cdot+\sqrt{\frac{\epsilon_0}{\mu_0}}(\ka_{DB})_{ik}{B}^k,\nonumber\\
     {M}_i&=&\frac{\partial \cl}{\partial B_i}+\frac{B_i}{\mu_0} =-\half(\ka_{HB})_{ik}\frac{{B}^k}{\mu_0}-\half\frac{{B}^j}{\mu_0}(\ka_{HB})_{ji}+\sqrt{\frac{\epsilon_0}{\mu_0}}{E}^k(\ka_{DB})_{ki},\label{pm00}
\end{eqnarray}
where $i, j$ and $k=1,2,3$.

Equation \ref{pm00} relates polarization $P$ and magnetization $M$ to the fields $E$ and $B$ and to SME coefficients which allow us to treat the Lorentz violating cosmos as a standard anisotropic optical medium. Actually, in the study of the optical linear response of materials to electromagnetic fields, equations such as \ref{pm00} are called constitutive relations and play a key role. These relations link the induced fields $D$ and $H$ to the applied fields $E$ and $B$ and they can be generalized to a $6\times 6$ matrix, as we will show below.

\section{Lorentz violation and optical analogy}\label{analogy}

As stated before, Lorentz symmetry violation in the SME framework can be compared to the propagation of photons in an anisotropic medium. This analogy allows for a deeper understanding of what the modified Maxwell equations in the SME framework predict as far as the propagation of photons is concerned.

\subsection{Lorentz violation in the SME framework}

Our starting point is the SME constitutive relations which define the fields $\textbf D$ and $\textbf H$ through a six-dimensional matrix equation whose elements can be deduced from equation \ref{pm00}. 

\beq
\left(
\begin{array}{c} 
\textbf {D} \\ \textbf {H} 
\end{array} 
\right)
=\hat{L}_V\left(
\begin{array}{c} 
\textbf E \\ \textbf B 
\end{array} 
\right)=
\left(
\begin{array}{cc} 
 1+\ka_{DE} & \ka_{DB} \\
 \ka_{HE} & 1+\ka_{HB} 
\end{array}
\right)
\left(
\begin{array}{c} 
\textbf E \\ \textbf B 
\end{array} 
\right),
\label{DH1}
\eeq

In general, the $\hat{L}_V$ matrix can be diagonalized, but this does not always have a physical meaning because we cannot accept a different reference frame for the field $\vec E$ and the field $\vec B$.

In any case, one can diagonalize the $\kappa_{DE}$ or $\kappa_{BH}$ submatrices and thus define a special reference frame $V_L$, which gives the simplest form for the $\hat{L}_V$  matrix. This reference frame is an absolute one, and the coefficients of $\hat{L}_V$ are fundamental components. 
\textcolor{black}{For example, if one diagonalizes the $\kappa_{DE}$ submatrice, t}he $\hat{L}_V$ in the \textcolor{black}{related} $V_L$ reference frame can be written as

\begin{align*}
 \hat{L}_V &= \left(\begin{array}{ccc|ccc}
    \mathcal{O}_{11}  & 0 & 0     &    \mathcal{O}_{14} &  \mathcal{O}_{15} &  \mathcal{O}_{16}  \\
 0 & \mathcal{O}_{22} & 0 & \mathcal{O}_{24} &  \mathcal{O}_{25} &  \mathcal{O}_{26}\\
 0&0&\mathcal{O}_{33}   &     \mathcal{O}_{34} &  \mathcal{O}_{35} &  \mathcal{O}_{36} \\
                   \hline 
                      \mathcal{O}_{41} &  \mathcal{O}_{42} &  \mathcal{O}_{43} & \mathcal{O}_{44} &  \mathcal{O}_{45} &  \mathcal{O}_{46} \\
  \mathcal{O}_{51} &  \mathcal{O}_{52} &  \mathcal{O}_{53} & \mathcal{O}_{54} &  \mathcal{O}_{55} &  \mathcal{O}_{56} \\  \mathcal{O}_{61} &  \mathcal{O}_{62} &  \mathcal{O}_{63} & \mathcal{O}_{64} &  \mathcal{O}_{65} &  \mathcal{O}_{66} \\
    \end{array}\right)
\end{align*}
 
\subsection{Magneto-electric medium} 
\label{me}

In the study of the linear response of materials to electromagnetic fields, constitutive relations can also be generalized to a $6\times 6$ matrix
\beq
\left(
\begin{array}{c} 
\textbf {D} \\ \textbf {H} 
\end{array} 
\right)
=
\left(
\begin{array}{cc} 
 \epsilon &\chi \\
 \xi &\mu^{-1}
\end{array}
\right)
\left(
\begin{array}{c} 
\textbf E \\ \textbf B 
\end{array} 
\right) ,
\label{DH}
\eeq
where in this equation, the 
matrix is divided into four $3\times 3$ blocks. The Electric Permittivity ($\epsilon$) as a tensor that describes the material’s electric response to the 
$E$-field and the inverse Magnetic Permeability ($\mu^{-1}$) as a tensor that models the material’s magnetic response to the $B$-field. Magnetoelectric Couplings ($\xi$ and $\chi$) are tensors that describe the interaction between the $E$ and $B$ fields,
\beq
\textbf{D}=\epsilon \textbf{E}+\chi \textbf{B},\,\,\,\,\,\,\,\,\,\,\,\,\textbf{H}=\xi \textbf{E}+\mu^{-1} \textbf{B}.
\eeq

Comparing the matrix equation for Lorentz-violating cosmos in the SME framework and the one for standard anisotropic media one clearly sees the equivalence between the $\kappa_{DE}$ and $\epsilon$ blocks, between $\kappa_{HB}$ and $\mu^{-1}$ blocks, between $\kappa_{DB}$ and $\chi$ blocks and between $\kappa_{HE}$ and $\xi$ blocks.

Let's note that in the standard textbook of electromagnetic media \cite{bornwolf}, $\mu^{-1}$ is assumed to be $\mu_0^{-1}\mathrm{Id}$ and $\xi=\chi=0$, so that Eq.\ref{DH} reduces to 
\beq
\textbf{D}=\epsilon \textbf{E},\,\,\,\,\,\,\,\,\,\,\,\,\textbf{H}=\mu_0^{-1} \textbf{B}.
\eeq

When the  matrix $\epsilon$ is real and diagonal but elements are not all equal, taking the following form  

\beq
\epsilon
=
\left(
\begin{array}{ccc} 
 \epsilon_x & 0 & 0 \\
 0 & \epsilon_y & 0\\
 0&0&\epsilon_z
\end{array}
\right)
,
\label{epsilon}
\eeq
linear birefringence occurs and a light linearly polarized becomes elliptically polarized. When $\epsilon_y = \epsilon_x$ the medium is said to be uniaxial \cite{bornwolf}. This anisotropy leads to two different refractive indices for polarizations perpendicular and parallel to the 
$z$-axis:
\beq
n_o=\sqrt{\epsilon_x\mu_0}, \,\,\,\,\,\,\, n_e=\sqrt{\epsilon_z\mu_0}, \,\,\,\,\,\,\,
\Delta=n_e-n_o.
\label{delta}
\eeq

When $\epsilon_x \neq \epsilon_y \neq \epsilon_z$ the medium is said to be biaxial  \cite{bornwolf}. Linear birefringence always occurs but the description of light propagation becomes more complex since the index of refraction is different for light polarization parallel to each axis \cite{Dreger1999}.

Obviously, the propagation of light in the general case of a magneto-electric crystal needs a more complex mathematical treatment \cite{Birss1967}, but again the medium behaves as a biaxial crystal showing linear magneto-electric birefringences as we show in the next paragraph in the case of a quantum vacuum in the presence of external electromagnetic fields.

\begin{figure*}[t!]
	\centering
	\centering
	\includegraphics[width=0.5\linewidth]{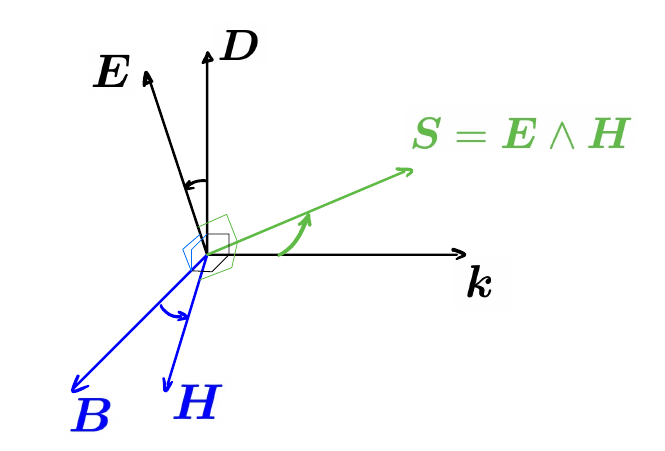}
	\caption{ The Poynting vector $\textbf S$  does not align with the wave vector $\textbf k$, indicating that the energy flow and phase velocity are in different directions. This phenomenon occurs in biaxial crystals,  and magneto-optical media. }
	\label{Wein}
\end{figure*}

Let's recall that in anisotropic media following the maxwell equation,
 \begin{eqnarray}
     \nabla\times {\textbf H}&=&\dot{ \textbf D }\rightarrow \textbf{k}\times {\textbf H}=  -\omega {\textbf D} \nonumber\\
     \nabla\times \textbf E&=&\dot{\textbf B }\rightarrow \textbf{k}\times {\textbf E}= - \omega {\textbf B}
     \label{max0}
 \end{eqnarray}
 therefore $\textbf D\perp \textbf H, \textbf k$ and $\textbf B \perp \textbf E, \textbf k$.  
 In such media, $\textbf D\nparallel E$ and $\textbf{B}\nparallel\textbf{H}$ and the Poynting vector ($\textbf S=\textbf E\times \textbf H$) may not be aligned with the wave vector ($\textbf k \nparallel \textbf S$), meaning that the phase velocity and energy flow in different directions (See Fig.~\ref{Wein}). This phenomenon occurs in uniaxial and biaxial crystals, metamaterials, and magneto-optical media. 

Moreover, such media permits only two monochromatic plane waves with two different linear polarizations and two different velocity to propagate in any given direction. These two polarizations corresponding to a given direction of propagation $\textbf k$ (See Fig.~\ref{Wein}) are perpendicular to each other \cite{bornwolf}. Now, since $\textbf S$ is different from $\textbf k$ and since $\textbf S$ depends on $\textbf E$ i.e. light polarization, orthogonal eigen-state polarizations separate each other (see e.g. \cite{Dreger1999}) unless $\textbf k$ is along some special direction that is related to the crystal axis. For example, in the case of a uniaxial crystal defined by the matrix \ref{epsilon} with $\epsilon_y = \epsilon_x$, $\textbf k$ must be perpendicular to the $z$ axis. Only in this case can the ellipticity induced by the linear birefringence be observed.

\subsection{Quantum Vacuum as a Magneto-Electric  Medium}
\label{QV}
To exemplify what kind of optical effects may exist in a Magneto-Electric medium and what elements of the $6x6$ matrix corresponding to the constitutive relations are related to the different effects let's study the quantum vacuum in the presence of external electromagnetic fields. In classical electrodynamics the vacuum is considered a passive and linear medium where electromagnetic waves propagate without alteration. However, within the framework of quantum field theory (QFT), the quantum vacuum exhibits nontrivial optical properties due to vacuum fluctuations and virtual particle interactions. This leads to a nonlinear response to external electromagnetic fields, a phenomenon primarily described by QED \cite{Battesti2013}.

The key manifestations of this nonlinearity are vacuum birefringences, where the polarization of light changes as it propagates through a strong electromagnetic field. These effects arise due to virtual electron-positron pairs interacting with the external field, modifying the permittivity and permeability of the vacuum so that it behaves as a nonlinear optical medium with properties analogous to standard materials.

Mathematically, the Euler-Heisenberg Lagrangian which is the effective Lagrangian of QED includes higher-order terms beyond Maxwell’s Lagrangian, 

The general expression can be written as \cite{Battesti2013}
\begin{equation}\label{genform}
    L= \sum_{i=0}^\infty \sum_{j=0}^\infty c_{i,j} F^i
    G^{j}.
\end{equation}
where 
\begin{equation}
F =  \left(\epsilon_{0}E^2 - {B^2 \over \mu_{0}}\right)
,\,\,\, G =\sqrt{\epsilon_{0} \over \mu_{0}} (\textbf{E} \cdot \textbf{B}).
\label{3}
\end{equation}

The lowest order terms gives the classical Maxwell Lagrangian $L_0=\frac{1}{2}F$, thus $c_{0,0}=0$, $c_{1,0}=\frac{1}{2}$ and  $c_{0,1}=0$.

The constitutive relations can be calculated as follows:
\begin{eqnarray}
\textbf{P} &=& \frac{\partial\mathcal{L}}{\partial\textbf{E}} - \epsilon_0 \textbf{E}, \nonumber\\
\textbf{M} &=& \frac{\partial\mathcal{L}}{\partial\textbf{B}} + \frac{\textbf{B}}{\mu_0}\label{Eq:H}.
\end{eqnarray}

In the presence of external static electromagnetic fields, the $\vec P$ and $\vec M$ are given by \cite{Battesti2013},

 \begin{eqnarray}\label{PolBir}
\textbf{P} &=& 4c_{2,0} \epsilon_{0}(\textbf{E}_{\omega}+\textbf{E}_0)\left(\epsilon_{0}E^2_0 - {B^2_0 \over \mu_{0}}+2\epsilon_0\textbf{E}_{\omega} \cdot \textbf{E}_0-{2 \textbf{B}_{\omega} \cdot \textbf{B}_0 \over \mu_0}\right)
 +\\
\nonumber  & & 2c_{0,2} {\epsilon_{0} \over \mu_{0}} (\textbf{B}_\omega+\textbf{B}_0)(\textbf{E}_\omega \cdot \textbf{B}_0+\textbf{E}_0 \cdot \textbf{B}_\omega+\textbf{E}_0 \cdot \textbf{B}_0),
\end{eqnarray}

\begin{eqnarray}\label{MagBir}
\textbf{M} &=& -4c_{2,0} {(\textbf{B}_{\omega}+\textbf{B}_0) \over \mu_0}\left(\epsilon_{0}E^2_0 - {B^2_0 \over \mu_{0}}+2\epsilon_0\textbf{E}_{\omega} \cdot \textbf{E}_0-{2 \textbf{B}_{\omega} \cdot \textbf{B}_0 \over \mu_0}\right)
 +\\
\nonumber  & & 2c_{0,2} {\epsilon_{0} \over \mu_{0}} (\textbf{E}_\omega+\textbf{E}_0)(\textbf{E}_\omega \cdot \textbf{B}_0+\textbf{E}_0 \cdot \textbf{B}_\omega+\textbf{E}_0 \cdot \textbf{B}_0),
\end{eqnarray}

where we have written the total electromagnetic fields $\textbf{E}$ and $\textbf{B}$ as the sum of the fields associated to the propagating waves $\textbf{E}_\omega$ and $\textbf{B}_\omega$ and the static ones, assuming that $E_\omega << E_0$ and $B_\omega << B_0$. We assume that the static fields are also homogeneous. 

The constitutive relations can therefore be written as 
\beq
\left(
\begin{array}{c} 
\textbf {P} \\ \textbf {M} 
\end{array} 
\right)
=
 \mathcal{O}
_{6\times 6}
\left(
\begin{array}{c} 
\textbf E \\ \textbf B 
\end{array} 
\right) ,
\label{DH}
\eeq
where $\mathcal{O}$ is a $6\times 6 $ tensor and determines the coupling between electric and magnetic fields as for a standard magneto-electric crystal.

We will assume in the following that the static electric field is directed along the $\hat{x}$ direction ($ \textbf{E}_0=E_{0}\hat i$), while the static magnetic field points in the direction perpendicular to $\hat{z}$, $ \textbf{B}_{0z}=0$. \textcolor{black}{This assumption is the most general one.}

Following constitutive relations $\mathcal{O}$ can be written as:
\begin{eqnarray}
 \mathcal{O}
=
\begin{pmatrix}
    \mathcal{O}_{11} & \mathcal{O}_{12} & 0 & \mathcal{O}_{14} & \mathcal{O}_{15} & 0
    \\  \mathcal{O}_{12} & \mathcal{O}_{22} & 0 & \mathcal{O}_{24} & \mathcal{O}_{25} & 0
    \\  0 & 0 &  \mathcal{O}_{33} &  0 & 0 & \mathcal{O}_{36} 
    \\  \mathcal{O}_{14} & \mathcal{O}_{24} & 0 & \mathcal{O}_{44} & \mathcal{O}_{45} & 0
    \\  \mathcal{O}_{15} & \mathcal{O}_{25} & 0 & \mathcal{O}_{45} & \mathcal{O}_{55} & 0
     \\  0 & 0 &  \mathcal{O}_{36} &  0 & 0 & \mathcal{O}_{66} 
\end{pmatrix}
 .
\label{matrix}
\end{eqnarray}
The expression of matrix elements in terms of fields and Lagrangian coefficients can be found in the Appendix \ref{appe}.

The matrix $\mathcal{O}$ can be separated into different matrices, each corresponding to a particular birefringence effect,
\begin{eqnarray}
     \mathcal{O}= \mathcal{O}_k+ \mathcal{O}_{CM}+ \mathcal{O}_j+ \mathcal{O}_{ME},
\end{eqnarray}
respectively, \textcolor{black}{Kerr \cite{Kerr1875}, Cotton-Mouton \cite{Cotton}, Jones \cite{Roth2000}, and Magneto-Electric \cite{Roth2002} birefringences (see also \cite{Battesti2013}).}
$\mathcal{O}_k$ corresponds to the effect of the presence of the $\vec E_0$ field,
\begin{eqnarray}
    \mathcal{O}_k=
    \begin{pmatrix}
   {O}_k^{11} & 0 & 0 & 0 & 0
    \\  0 &{O}_k^{22} & 0 & 0 & 0 & 0
    \\  0 & 0 &  {O}_k^{33} &  0 & 0 & 0
    \\ 0 & 0 & 0 & {O}_k^{44}& 0 & 0
    \\  0 & 0 & 0 & 0 & {O}_k^{55} & 0
     \\  0 & 0 &  0 &  0 & 0 & {O}_k^{66} 
\end{pmatrix}
 .
\label{Ok}
\end{eqnarray}
$ \mathcal{O}_k$ is diagonal birefringence axis are parallel and perpendicular to $\vec E_0$,  therefor along $\hat z$ and $\hat y$ axis.
The values of the elements can be found in Appendix A when $\vec B_0 = 0$.

The matrix $\mathcal{O}_{CM}$ which corresponds to the Cotton-Mouton effect is given as,
\begin{eqnarray}
 \mathcal{O}_{CM}
=
\begin{pmatrix}
    \mathcal{O}^{CM}_{11} & \mathcal{O}^{CM}_{12} & 0 & 0 &0 & 0
    \\  \mathcal{O}^{CM}_{12} & \mathcal{O}^{CM}_{22} & 0 &0 & 0 & 0
    \\  0 & 0 &  \mathcal{O}^{CM}_{33} &  0 & 0 & 0
    \\ 0 & 0 & 0 & \mathcal{O}^{CM}_{44} & \mathcal{O}^{CM}_{45} & 0
    \\  0 & 0& 0 & \mathcal{O}^{CM}_{45} & \mathcal{O}^{CM}_{55} & 0
     \\  0 & 0 & 0 &  0 & 0 & \mathcal{O}^{CM}_{66} 
\end{pmatrix}
 .
\label{CM}
\end{eqnarray}
The values of the elements can be found in Appendix A when $E_0 = 0$.
The matrix $ \mathcal{O}_{CM}$ is not diagonal because the birefringence axis are parallel and perpendicular to $\vec B_0$, which is neither aligned with the $\hat z$ axis nor along the  $\hat y$ axis. 
One could diagonalize $\mathcal{O}_{CM}$, but obviously $ \mathcal{O}_k$ could become non-diagonal.

The matrix $\mathcal{O}_{J}$, which corresponds to the Jones birefringence due to the presence of $E_0$ and $B_{0x}$, is given by,
\begin{eqnarray}
 \mathcal{O}_J
=
\begin{pmatrix}
    0& 0 & 0 & \mathcal{O}^{J}_{14}  &0 & 0
    \\0& 0 & 0 &0&  \mathcal{O}^{J}_{25} &  0
    \\ 0& 0 & 0 &0& 0& \mathcal{O}^{J}_{36} 
    \\  \mathcal{O}^{J}_{41} &0& 0 & 0 &0&0 
    \\  0 &\mathcal{O}^{J}_{42}& 0& 0 & 0 & 0
     \\ 0 & 0 &  \mathcal{O}^{J}_{66} & 0 & 0 &  0 
\end{pmatrix}
 .
\label{CM}
\end{eqnarray}

$\mathcal{O}_{ME}$, corresponding to the presence of $E_0$ and $B_{0y}$, is given by,
\begin{eqnarray}
 \mathcal{O}_{ME}
=
\begin{pmatrix}
    0& 0 & 0 &0 & \mathcal{O}^{ME}_{15}  & 0
    \\0& 0 & 0 &  \mathcal{O}^{ME}_{24}& 0&  0
    \\ 0& 0 & 0 &0& 0&0
    \\0&  \mathcal{O}^{ME}_{41} &0&  0 &0&0 
    \\  \mathcal{O}^{ME}_{51}& 0 &0& 0 & 0 & 0
     \\ 0 & 0 &  0& 0 & 0 &  0 
\end{pmatrix}
 .
\label{CM}
\end{eqnarray}
Both Jones and $ME$ birefringences have the particularity that axis are oriented at $45^\circ$ from the $E_0$ field. As a result, light polarized at $+45^\circ$ or $-45^\circ$ experiences opposite refractive indices ($n^+=n^-$). This means that these types of birefringence are directional and they change sign when the propagation vector changes sign thus they cannot be observed in the linear optical cavity, unlike the Kerr and $CM$ linear birefringence, which can be.

Studying the Quantum vacuum case we show that each element of the constitutive equation matrix corresponds to a particular birefringence which can be directional or not, for example. Because of the perfect analogy between SME theory and crystal optics, these general properties of the matrix $\mathcal{O}$ apply to the case of a Lorentz violating cosmos.

\subsection{Calculation of birefringence in the Presence of LV}
\label{sec3}

We can also recover this analogy  between anisotropic optics and SME Lorentz-violation, by using the expressions for polarization and magnetization in a system governed by Lorentz-violating photon Lagrangian Eq.~(\ref{lagrangian}). For the sake of simplicity, we assume that the matrix $\mathcal{O}$ is diagonal both for the $\kappa_{DE}$ and $\kappa_{BH}$ sub-matrices like in the case of a biaxial magneto-electric crystal.

To begin, we start from the modified Maxwell equations, which describe the evolution of the electric and magnetic fields in the presence of Lorentz violation. These equations are written as:
\begin{eqnarray}
   \nabla\cdot{\textbf D}&=&0 \,\,\,\,\,\,\,\,\,\,,\,\,\,\,\,\,\,\,\,\,\,\,\,\,\,\,\,\,\,\,\,\,\nabla\times {\textbf H}-\frac{\partial \textbf D }{\partial t}=0\nonumber\\
     \nabla\cdot{\textbf B}&=&0 \,\,\,\,\,\,\,\,\,\,,\,\,\,\,\,\,\,\,\,\,\,\,\,\,\,\,\,\,\,\,\,\,\nabla\times \textbf E+\frac{\partial \textbf B }{\partial t}=0\label{max}
\end{eqnarray}
 The terms introduced by Lorentz violation modify the relations between these fields, leading to new coupling constants and interactions that can be used to derive expressions for polarization ($\textbf{P}$) and magnetization ($\textbf{M}$) which  are typically defined in terms  of the electric displacement 
$\textbf{D}$ and the magnetic field 
$\textbf{H}$ as follows:
\begin{eqnarray}
    \textbf{D}&=&\epsilon_0\textbf{E}+\textbf{P}, \,\,\,\,\,\,\,\,\,\,\,\,\,\,
    \textbf{H}=\frac{\textbf{B}}{\mu_0}-\textbf{M},
    \label{dh}
\end{eqnarray}
these equations show how the fields are modified by the medium's response to external electric and magnetic fields. By substituting Eq.~(\ref{dh}) in (\ref{max}) the maxwell equation can be written by
\begin{eqnarray}
 \nabla\cdot{\textbf D}&=&0 \,\,\,\,\,\,\,\,\,\,,\,\,\,\,\,\,\,\,\,\,\,\,\,\,\,\,\,\,\,\,\,\,\frac{1}{\mu_0}\nabla\times {\textbf B}-\nabla\times {\textbf M}-\epsilon_0\frac{\partial \textbf E }{\partial t}-\frac{\partial \textbf P }{\partial t}=0\nonumber\\
     \nabla\cdot{\textbf B}&=&0 \,\,\,\,\,\,\,\,\,\,,\,\,\,\,\,\,\,\,\,\,\,\,\,\,\,\,\,\,\,\,\,\,\nabla\times \textbf E+\frac{\partial \textbf B }{\partial t}=0\label{max}   
\end{eqnarray}
Once the expressions for polarization and magnetization are derived, they provide crucial information about how electromagnetic waves propagate in such a medium. 

In this case, we will consider a wave propagating in this medium, assuming that the direction of the photons is along the $z$-axis. The electric field associated with the propagating wave with reflective index $n$ is given as:
\begin{eqnarray}
    \textbf{E}_W=\textbf{E}_0e^{iw(\frac{n}{c}{z}-t)}\label{wave}.
\end{eqnarray}

 Now by inserting the Eqs.~({\ref{wave} and \ref{pm00}}) into the Maxwell equations (\ref{max}), one gets in the polarization plane $(x,y)$   
\begin{equation}
\begin{pmatrix}
-n^2(1+\ka_{HB}^{11})+2+\ka_{DE}^{11}& n(\ka_{DB}^{21}-\ka_{DB}^{12}) \\
n(\ka_{DB}^{21}-\ka_{DB}^{12}) & -n^2(1+\ka_{HB}^{22})+2+\ka_{DE}^{22}
\end{pmatrix}
\textbf{E}_W=\textbf{E}_W.
\end{equation}
In this case, the eigenmodes corresponding index of refraction are:
\begin{eqnarray}
n_{x} = 1 +\half ( \ka_{DE}^{11}-\ka_{HB}^{11}),\nonumber\\
n_{y} = 1 +\half (\ka_{DE}^{22}-\ka_{HB}^{22}),
\label{nxny}
\end{eqnarray}
where $n_{x}$ and $n_{y}$are the index of refraction for light polarized in the plane perpendicular to the wave directions.

The anisotropy $\Delta n$ is equal to:
\begin{equation}\label{Deltan}
\Delta n_\mathrm{CM} = n_{x} - n_{y} = \frac{1}{2}( \ka_{DE}^{11}- \ka_{DE}^{22})+ \frac{1}{2}( \ka_{HB}^{22}- \ka_{HB}^{11}),
\end{equation}

which is the expected linear combination of $\mathcal{O}$ non zero elements, showing once more the SME Lorentz violating cosmos behaves as a standard magneto-electric medium (see e.g. eq. \ref{delta}).

\subsection{Optical activity}
\label{oa}

In the previous sections, we have just dealt with real elements for $\mathcal{O}$. As shown, these real elements are related to linear birefringences. Polarized light propagating in a linear birefringent medium acquires an ellipticity. The major axis of the ellipse described by the light $E$ field vector is aligned with the original polarization direction.

\textcolor{black}{Actually, in the framework of cosmic Lorentz violation has been largely discussed what is called cosmic birefringence (see e.g. \cite{Komatsu}) i.e. a rotation of the light polarization direction. As far as we understand, this is the effect on linearly polarized light propagating in the cosmos that attracted most the attention. From our approach based on the optical point of view a medium having the property of rotate light polarization is called an optical active medium. Optical activity is related to a circular birefringence which means that light velocity depends on the circular mode of the light polarization, clockwise or anticlockwise.}

From optics theory, we know that optical activity causing the rotation of polarization is present when $\textbf D$ vector can be expressed as

\begin{eqnarray}
    \textbf{D}=\epsilon_0\textbf{E}+ g \nabla\times {\textbf E},
    \label{optict0}
\end{eqnarray}

which give non-diagonal imaginary terms in the matrix $\mathcal{O}_{OA}$ block giving $\textbf{D}=\mathcal{O}_{OA}\textbf{E}$, with

\begin{eqnarray}
\begin{pmatrix}
\textbf{D} \\
\textbf{H}
\end{pmatrix}
=\mathcal{O}_{OA}
\begin{pmatrix}
\textbf{E} \\
\textbf{B}
\end{pmatrix}.
\label{opty2}
\end{eqnarray}

Now $\nabla\times {\textbf E}=\frac{\partial}{\partial t}\textbf{ B}=i\omega\textbf{B}$ which also means that optical activity can be related to imaginary diagonal elements of $\kappa _{DB}$. 

Consequently, imaginary non-diagonal elements in the matrix $\mathcal{O}_{OA}$ with $\textbf{H}=\mathcal{O}_{OA}\textbf{B}$ and imaginary diagonal elements in the block of the $\mathbf{O}_{OA}$ matrix giving $\textbf{H}$ as a function of $\textbf{E}$ also are associated to optical activity. 

The $\hat{L}_V $ can contain both real and imaginary components. Because of the perfect analogy between $\hat{L}_V $ and $\mathbf{O}_{OA}$, it is clear that if the elements of $\hat{L}_V $ are imaginary then the circular birefringence is expected. This corresponds to a cosmic optical activity, which results in the rotation of the polarization vector for light propagating in a Lorentz violating cosmos.

\subsection{Dichroism and apparent polarization rotation }
 \label{dich}

Together with ellipticity and rotation dichroism is an important effect which corresponds in a change of the polarization of light propagating in a medium. In general terms, one can talk of dichroism when optical losses depend on light polarization. Losses can be due to polarization selective absorption or diffusion or deviation that reduce the contribution of light polarized in a certain direction to the total polarization of the original propagating light. Dichroism, which is therefore a losses of light energy, is typically observed as an apparent rotation of light polarization. In fact, since one component of the original light polarization vector is observed, the resulting light polarization vector seems to turn towards the remaining component.

The distinction between dichroism (energy walk-off) and birefringence (phase delay) is critical in Lorentz-violating systems. Following our previous example, for a wave propagating through the Lorentz-violating vacuum over a distance $L$, the phase difference accumulates as:

\begin{eqnarray}
\Delta \phi = \frac{2\pi L}{\lambda} \left(n_x - n_y\right),
\end{eqnarray}

where $n_x$ and $n_y $ are the real refractive indices for the  $x$-  and $y$-polarized modes in Eq.~(\ref{nxny}), and $ \lambda $ is the wavelength. In fact, $\Delta \phi$ is the temporal delay between polarizations (birefringence).

As already shown, the effective medium arising from Lorentz-violating coefficients of the extended Standard Model, the Poynting vector and the wave vector are not aligned. This misalignment implies that the direction of energy flow differs from the direction of wave propagation. Such a feature in an anisotropic medium can lead to polarization-dependent absorption or transmission, even in the absence of imaginary Lorentz-violating terms. In other words, this difference between the directions of $\vec S$ and $\vec k$ may result in an effective dichroism, originating from the geometric and anisotropic structure of the medium rather than from conventional loss mechanisms. This form of dichroism can be viewed as a physical consequence of the CPT-even Lorentz-violating sector of the Lagrangian.

Dichroism is caused by the fact that the energy flux of the two polarizations spatially separates, leading  \textit{lateral displacement} ($\Delta x$) which is in principle measurable. Following the classical optics theory \cite{bornwolf}, $\Delta x$ is proportional to the linear birefringence i.e. to a linear combination of the SME parameters, as in our previous example (eq. \ref{Deltan}). 

The apparent polarization rotation related to this cosmic dichroism is also in principle measurable. For long distance of propagation one may assume that the original light beam splits completely in two and light observed on earth is polarized along one of the two polarization eigen-values.

\section{Conclusions}\label{secconc}

The analogy between the SME Lorentz-violating cosmos and crystal optics has observational consequences that should be studied in details, but such a study is largely beyond the scope of this paper which is mainly dedicated to explain this analogy more extensively than originally done in ref. \cite{Colladay1998}. In the following therefore will we restrain ourselves to discuss only some general fact.

We have clearly show that in the Magneto-Electric medium in the presence of external fields in a Lorentz violation cosmos linear birefringence appears and that each of them are related to the group of elements in the general $\hat{L}_V $ matrix describing the Lorentz violation in the cosmos and that can be written in the absolute reference frame i.e. the one that reduces the numbers of elements of $\hat{L}_V $ non zero by diagonalizing at least one of the sectors.

On the other hand, the components of the polarization vector related to $\mathcal{V}_L$ of the propagating light, for example in an optical cavity, as observed on the Earth, change over time due to its motion relative to the $\mathcal{V}_L$ frame.

Thus, the components of the $\hat k$ vector can be written as 
\begin{eqnarray}
\hat{k}=
\begin{pmatrix}
 \sin\theta\cos\phi \\
 \sin\theta\sin\phi\\
 \cos\theta
\end{pmatrix}
\end{eqnarray}
with $|\hat k|^2=1$, where the angle are defined as usual in the spherical set of coordinates, and 
\begin{eqnarray}
\textbf{E}=
\begin{pmatrix}
E \sin\theta_E\cos\phi_E \\
 E\sin\theta_E\sin\phi_E\\
E \cos\theta_E
\end{pmatrix}
,
\end{eqnarray}
Then,  $\textbf{B}$ can be calculated by using the equation $\textbf{B}\propto \textbf{k}\times\textbf{E}$.
Thus for linear polarized light, circular polarized light can be written following standard optical rules.

One sees therefore that any terrestrial experiment result depends on four angular parameters describing the polarization direction and the wave vector with respect to the absolute Lorentz-violating $\mathcal{V}_L$ frame. This unknown frame should be determined by observing the variation of the experimental results as a function of time and therefore as a function of the four angular parameters. 

To exemplify all that, let's imagine a simple uniaxial cosmos with $\mathbf{O}_{11}=\mathbf{O}_{22}=\epsilon_0$, $\mathbf{O}_{33}\neq\epsilon_0$, $\mathbf{O}_{44}=\mathbf{O}_{55}=\mathbf{O}_{66}=1/\mu_0$ and all the other elements equal to zero. If in an experiment light is polarized along $Z$ axis but the apparatus moves only in the $xy$ plane, no effect can be detected even if the cosmos is Lorentz violating.

Let's finally consider also the case of the observation from earth of a radiation coming from all the directions in a SME Lorentz violating cosmos. Let's also assume that the original polarization pathern of the radiation is known as in the case of CMB \cite{Komatsu}. This background is observed as if the earth were surrounded by a magneto-electric crystal. Light coming from one direction has its own velocity different from other directions. Most of all directions would be affected by dichroism since original light is split in the eigen-vector polarizations depending on the direction and therefore polarization should apparently rotate. In principle studying the change in polarization with respect to the direction could permits to characterize the $\mathcal{V}_L$ frame or extrude its existence. 

In conclusion, research into Lorentz symmetry breaking using the SME framework connects theoretical predictions with experimental observations. By studying how photon propagation is affected, this work has the potential to uncover new principles of physics and expand our understanding of the universe beyond the Standard Model.

We have clearly show that, as in the Magneto-Electric medium in the presence of external fields, in a Lorentz violation cosmos linear birefringence appears, but also light split following its polarization with respect to the axis of the absolute frame related to the Lorentz violation. This gives an apparent rotation of the polarization plane. Rotation of the polarization also appears as an optical activity if SME coefficients are imaginary.

Eventually, existing limits on the SME coefficients should be maybe reformulated having in mind the analogy described in this paper and hopefully novel experimental tests being put forward opening new windows on our universe.

\appendix
\section{Matrix elements}\label{appe}
   \begin{eqnarray}
    \mathcal{O}_{11}&=& 12 c_{2,0} \epsilon_{0}^2E_0^2-4c_{2,0} {\epsilon_{0} \over \mu_{0}}B_0^2+2c_{0,2} {\epsilon_{0} \over \mu_{0}}B_{0x}^2,\nonumber\\
    \mathcal{O}_{12}&=&\mathcal{O}_{21}= 2c_{0,2} {\epsilon_{0} \over \mu_{0}}B_{0x}B_{0y},\nonumber\\
    \mathcal{O}_{14}&=& \mathcal{O}_{41}=- 8 c_{2,0} {\epsilon_{0} \over \mu_{0}}E_0B_{0x}+ 2c_{0,2} {\epsilon_{0} \over \mu_{0}}E_0B_{0x},\nonumber\\
 \mathcal{O}_{15}&=&\mathcal{O}_{51}=- 8 c_{2,0} {\epsilon_{0} \over \mu_{0}}E_0B_{0y},\nonumber\\
 \mathcal{O}_{22}&=& 4 c_{2,0} (\epsilon_{0}^2E_0^2- {\epsilon_{0} \over \mu_{0}}B_0^2)+2c_{0,2} {\epsilon_{0} \over \mu_{0}}B_{0y}^2,\nonumber\\
 \mathcal{O}_{24}&=&\mathcal{O}_{42}= 2c_{0,2} {\epsilon_{0} \over \mu_{0}}E_{0}B_{0y},\nonumber\\
 \mathcal{O}_{25}&=&\mathcal{O}_{52}=\mathcal{O}_{36}=\mathcal{O}_{63}= 2c_{0,2} {\epsilon_{0} \over \mu_{0}}E_{0}B_{0x},\nonumber\\
 \mathcal{O}_{33}&=& 4 c_{2,0}( \epsilon_{0}^2E_0^2-{\epsilon_{0} \over \mu_{0}}B_0^2),\nonumber\\
 \mathcal{O}_{44}&=& -4 c_{2,0} ({\epsilon_{0} \over \mu_{0}}E_0^2- {B_0^2 \over \mu_{0}^2})+8 c_{2,0} {B_{0x}^2 \over \mu_{0}^2}+2c_{0,2} {\epsilon_{0} \over \mu_{0}}E_{0}^2,\nonumber\\
  \mathcal{O}_{45}&=&\mathcal{O}_{54}= 8c_{2,0} {B_{0x}B_{0y} \over \mu_{0}^2},\nonumber\\
  \mathcal{O}_{55}&=& -4 c_{2,0} ({\epsilon_{0} \over \mu_{0}}E_0^2- {B_0^2 \over \mu_{0}^2})+8 c_{2,0} {B_{0y}^2 \over \mu_{0}^2},\nonumber\\
   \mathcal{O}_{66}&=& -4 c_{2,0} ({\epsilon_{0} \over \mu_{0}}E_0^2- {B_0^2 \over \mu_{0}^2})\label{elements}
\end{eqnarray}

\end{document}